# Ferromagnetic Epitaxial μ-Fe$_2$O$_3$ on β-Ga$_2$O$_3$:
# A New Monoclinic form of Fe$_2$O$_3$

*John S. Jamison[1], Brelon J. May[1], Julia I. Deitz [1,2], Szu-Chia Chien[1], David W. McComb[1,2], Tyler J. Grassman[1,3], Wolfgang Windl[1], and Roberto C. Myers[1,3]*

[1]Department of Materials Science and Engineering, The Ohio State University, 43210

[2]Center for Electron Microscopy and Analysis, The Ohio State University, 43212

[3]Department of Electrical and Computer Engineering, The Ohio State University, 43210

Abstract: Here we demonstrate a new monoclinic iron oxide phase (μ-Fe$_2$O$_3$), epitaxially stabilized by growth on (010) β-Ga$_2$O$_3$. Density functional theory (DFT) calculations find that the lattice parameters of freestanding μ-Fe$_2$O$_3$ are within ~1% of those of β-Ga$_2$O$_3$ and that its energy of formation is comparable to that of naturally abundant Fe$_2$O$_3$ polytypes. A superlattice of μ-Fe$_2$O$_3$/β-Ga$_2$O$_3$ is grown by plasma assisted molecular beam epitaxy, with resulting high-resolution x-ray diffraction (XRD) measurements indicating that the μ-Fe$_2$O$_3$ layers are lattice-matched to the substrate. The measured out-of-plane (**b**) lattice parameter of 3.12 ± 0.4 Å is in agreement with the predicted lattice constants and atomic-resolution scanning transmission electron microscopy (STEM) images confirm complete registry of the μ-Fe$_2$O$_3$ layers with β-Ga$_2$O$_3$. Finally, DFT modeling predicts that bulk μ-Fe$_2$O$_3$ is antiferromagnetic, while the interface region between μ-Fe$_2$O$_3$ and β-Ga$_2$O$_3$ leads to ferromagnetic coupling between interface Fe$^{3+}$ cations selectively occupying tetrahedral positions. Magnetic hysteresis persisting to room temperature is observed via SQUID measurements, consistent with the computationally predicted interface magnetism.

Iron oxides contain two of the four most abundant elements on earth and constitute one of the most important material classes for society and industry with a wide variety of chemical, electronic, and magnetic properties besides being mined for iron production. Among the known iron oxides, $Fe_2O_3$ (or iron (III) oxide) is the most widely used compound as the source of industrial iron production. As such, it is extremely well studied; while high-pressure research has discovered a number of sub-stoichiometric iron oxides in the past decade[1], no new crystal structure for stoichiometric iron oxides has been brought forward in nearly a century. The two polytypes most common at standard pressure are the α (hematite) and γ (maghemite) phases which have the corundum (trigonal) and inverse-spinel (cubic) structures, respectively. The ε phase is a rare orthorhombic polytype which when found typically accompanies the α and γ phases. Additionally, there are several polytypes of $Fe_2O_3$ which occur at high temperatures and pressures that may exist naturally within the Earth's mantle. Finally, the β phase (bixbyite) which has a cubic structure occurs naturally as $FeMnO_3$ but can be synthesized as a pure iron (III) oxide by nanoparticle synthesis in an $SiO_2$ matrix[2] or grown epitaxially on a $Sn:In_2O_3$ (ITO) / yttria-stabilized zirconia (YSZ) heterostructure.[3]

Our discovery of a new monoclinic phase of $Fe_2O_3$ described in this letter starts with the beta phase of Gallium Oxide (β-$Ga_2O_3$), which has recently moved low-symmetry crystal structures into the center of attention due to its rivaling or even superior properties in high-power electronic devices in comparison to traditional high-symmetry materials.[4] β-$Ga_2O_3$ is appealing for the mass production of vertical wide band gap semiconductor heterostructures due to its very high breakdown voltage ($E_b = 8$ MV/cm) and availability of high quality bulk substrates.[5] Additionally, because β-$Ga_2O_3$ is composed of relatively light elements, electron spin lifetimes are long, on the order of 100 ns.[6] This, combined with high $E_b$, make it a promising candidate for spin-based field effect transistors (spin-FETs), provided magnetic doping and ferromagnetic ordering is possible. From this standpoint, $Fe^{3+}$ alloyed β-$Ga_2O_3$ presents a compelling and unexplored approach for developing magnetic semiconductors based on a single, simple alloy system. Furthermore, the low lattice thermal conductivity[7] of β-$Ga_2O_3$ is an attractive property in thermoelectrics and, in conjunction with magnetically ordered $Fe^{3+}$ dopants as a source for magnon-electron drag, may enable high thermoelectric efficiency.[8] This motivated are initial efforts to study the incorporation of Fe into the β-$Ga_2O_3$ lattice.

Once we found that Fe can be easily doped into the β-$Ga_2O_3$ lattice, the question arose if the monoclinic structure of β-$Ga_2O_3$ could offer a unique template on which to grow a new monoclinic phase of $Fe_2O_3$. Figure 1 shows that the (010) planes contain an equal number of $GaO_6$ octahedra and $GaO_4$ tetrahedra. $Fe^{3+}$ ions are larger than $Ga^{3+}$ ions, but only by ~4% in either the octahedral or tetrahedral coordination.[9] Therefore, it should be possible to epitaxially stabilize a phase of $Fe_2O_3$ isomorphic to that of β-$Ga_2O_3$, so long as its formation energy is reasonably close to those of the known iron oxides. In this study we first show by means of density-functional theory (DFT) calculations that this is indeed the case, and then demonstrate the growth of such a polytype of $Fe_2O_3$ (hereby referred to as μ-$Fe_2O_3$) via plasma assisted molecular beam epitaxy (PAMBE).

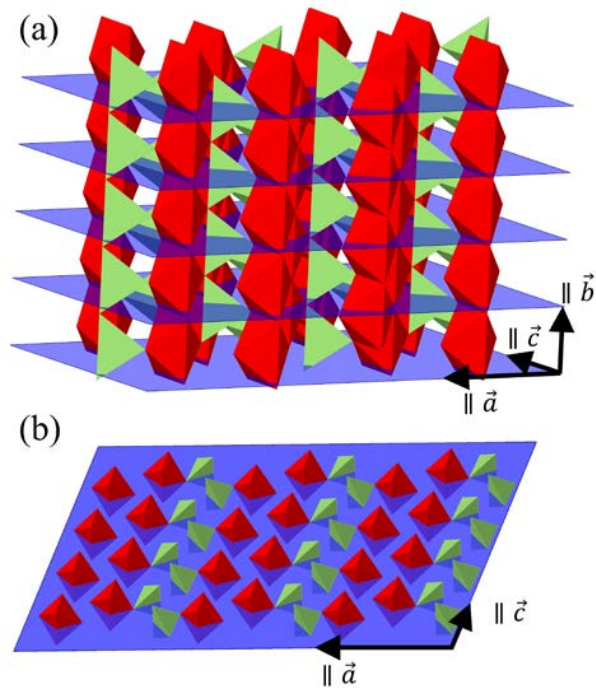

**Figure 1.** (a) Crystal structure of β-$Ga_2O_3$ with the (010) planes marked by the blue sheets. (b) A single (010) plane showing the equal number of $GaO_4$ tetrahedra (green) and $GaO_6$ octahedra (red). The lattice vectors are marked by the black arrows.

The total energy of μ-$Fe_2O_3$ in comparison to that of other known iron oxides was calculated within Dudarev's PBE+U approach to DFT[10] with an effective Hubbard U of 3 eV. In a recent comparative study,[11] it was shown that this combination results in excellent energetics and good structural parameters for iron oxides, whereas HSE with a mixing parameter larger than ~0.1, used in e.g. for $Ga_{2n-1}Fe_1O_{3n}$ was found to result in insufficient energetics[12]. Collinear spin-resolved calculations were performed in VASP using PAW potentials and a kinetic-energy cutoff of 400 eV. Brillouin zone sampling was done using a Monkhorst-Pack k-point mesh for k-space integration, with the product of the number of k-points and corresponding lattice vector equaling ~45 Å (60 Å for interface structures), and Blöchl's tetrahedron integration for the final energy.

$Fe_2O_3$ structures examined included the lowest-energy polymorphs, specifically α-, β-, γ-, ε-, and the new structure, μ-$Fe_2O_3$. Note that for γ-$Fe_2O_3$, which is a defective spinel structure, the energy-minimized vacancy arrangement from Ref. [13] was used. Inclusion of the recently proposed (and allegedly stable) monoclinic ς-structure,[14] which was reported to be created from a high-temperature phase transformation at 30 GPa from cubic β-$Fe_2O_3$, with $a$ = 9.68 Å, $b$ = 10.00 Å, $c$ = 8.95 Å, and β = 101° was also attempted. However, this phase was found to be thermodynamically unstable relative to β-$Fe_2O_3$. For both the ambient pressure and 30 GPa 4 relaxations, the three lattice constants end up with nearly equal values of 9.38 Å, and the monoclinic angle relaxed to 90° ± 2°. All other structures started from files downloaded from Materials Project.[15]

**TABLE 1**. Present DFT and experimental literature results for previously known polymorphs of $Fe_2O_3$ in comparison to theoretical predictions for μ-$Fe_2O_3$. The energy difference $\Delta E$ is calculated per $Fe_2O_3$ formula unit relative to the lowest-energy α-phase. Experimental values are from Springer Materials online[16-19]. μ-epi is epitaxially lattice matched to β-$Ga_2O_3$ with relaxed $b$ lattice parameter.

| Polymorph | Space group | Present DFT results | | | | Experiment | | |
|---|---|---|---|---|---|---|---|---|
| | | $\Delta E$ (eV) | $a$ (Å) | $b$ (Å) | $c$ (Å) | β (°) | $a$ (Å) | $b$ (Å) | $c$ (Å) |
| α[16] | $R\overline{3}c$ | 0.00 | 5.02 | | 13.73 | | 5.04 | | 13.73 |
| β[17] | $Ia\overline{3}$ | 0.20 | 9.38 | | | | 9.39 | | |
| γ[18] | $P4_12_12$ | 0.07 | 8.33 | | 24.93 | | 8.34 | | 25.02 |
| ε[19] | $Pna2_1$ | 0.10 | 5.08 | 8.77 | 9.47 | | 5.09 | 8.77 | 9.47 |
| μ | $C2/m$ | 0.14 | 12.42 | 3.07 | 5.88 | 103.9 | N/A | | |
| μ-epi | $C2/m$ | 0.17 | 12.23 | 3.16 | 5.80 | 103.7 | ? | 3.12±0.4 | ? |

Table 1 summarizes our results for freestanding (non-epitaxial) polymorphs of $Fe_2O_3$. All relaxed structures are found to exhibit antiferromagnetic ordering. For μ-$Fe_2O_3$, we find the lowest-energy magnetic structure to have opposite spins on octahedral and tetrahedral sites, both with a magnitude of 4 $\mu_B$. While there are competing antiferromagnetic arrangements with only slightly higher total energies, ferri- and ferromagnetic structures have considerably higher energies. Thus, any observed magnetic response must come from lattice imperfections, interface effects, or canting (as is the case in hematite). Overall, we find that μ-$Fe_2O_3$ has an energy per formula unit that is only 0.14 eV higher than the most stable α-phase, well within the range of 0.07, 0.10, and 0.20 eV for the known γ-, ε-, and β-phases, respectively, and thus energetically unproblematic. Furthermore, we find that its theoretical lattice constants are within 1% of those of β-$Ga_2O_3$. Consequently, the strain energy is only 0.03 eV per formula unit when μ-$Fe_2O_3$ is lattice matched to β-Ga$_2$O$_3$ while allowing out-of-plane relaxation. Recent work on the doping of Fe into Ga$_2$O$_3$ showed variation in lattice constant as a function of Fe content for (Ga$_{1-x}$Fe$_x$)$_2$O$_3$.[20] Linear extrapolation of the **b** lattice constant given in Ref. 20 from x = 0.3 match the predicted lattice constant of unstrained bulk μ-Fe$_2$O$_3$. Leftover strain in the sintered compounds may be why the **a** and **c** lattice constants do not match from linear extrapolation.[21]

Multilayer structures of μ-Fe$_2$O$_3$ / β-Ga$_2$O$_3$ are grown via PAMBE in a Riber/MBE Control M7 system equipped with Ga and Fe effusion cells and a Veeco oxygen plasma source. The growth conditions, unless otherwise stated, are identical to those in Ref. 22. The growth rate of Fe is calibrated at a beam equivalent pressure (BEP) 5.3×10$^{-8}$ torr by depositing Fe on a cold Si (111) wafer for 30 mins and measuring the film thickness using cross-sectional scanning electron microscopy (SEM). For simplicity, assuming the film to be fully dense body centered cubic (BCC) phase, the growth rate corresponds to an atomic flux of 2.8×10$^{16}$ cm$^{-2}$ min$^{-1}$. Using this value, an equivalent growth rate for the hypothetical μ-Fe$_2$O$_3$ phase is calculated to be 0.4 ML/s. The β-Ga$_2$O$_3$ growth rate is 0.5 ML/s at a Ga BEP of 8.6×10$^{-8}$ torr.[22]

The substrate is a 5 × 5 × 0.5 mm$^3$ unintentionally-doped (010) β-Ga$_2$O$_3$ single crystal (Tamura Corporation). Before growth, the substrate is first cleaned by sonication for 5 mins in acetone followed by the same sonication in methanol and then finally isopropanol. The substrate is indium bonded to the unpolished side of a 3-inch silicon wafer. Once in the chamber a final cleaning is performed by heating to 800 °C (measured by thermocouple) for 10 minutes while being exposed to the O$_2$ plasma. The forward power of the plasma is 300 W with ≤ 1 W reflected. During the pre-clean and growth, the flow of O$_2$ is continuously monitored and adjusted to maintain a growth chamber pressure of 1.5×10$^{-5}$ torr. Following the pre-cleaning procedure, the substrate is cooled to the growth temperature of 700 °C. The layer structure is designed to determine the thickness of

μ-Fe$_2$O$_3$ that can be grown on β-Ga$_2$O$_3$ (010) before the deterioration of a two-dimensional (2D) growth mode or clear phase change as observed by *in-situ* reflection high energy electron diffraction (RHEED). First, a 75-nm thick buffer layer of β-Ga$_2$O$_3$ is grown followed by a superlattice with varying thicknesses of μ-Fe$_2$O$_3$ (0.2-50 ML) with 15-nm thick β-Ga$_2$O$_3$ spacers, shown schematically in Figure 2a. A total of 33 periods are grown with the final layer being a 15-nm thick capping layer of β-Ga$_2$O$_3$. Throughout the entire growth (i.e. during both β-Ga$_2$O$_3$ and μ-Fe$_2$O$_3$ layers) an identical streaky RHEED pattern is observed indicative of the preservation of a 2D surface without an identifiable change in lattice parameter (Figure 2b,c). Even for the thickest layers of μ-Fe$_2$O$_3$ (50 ML) no clear change in the RHEED pattern is observed.

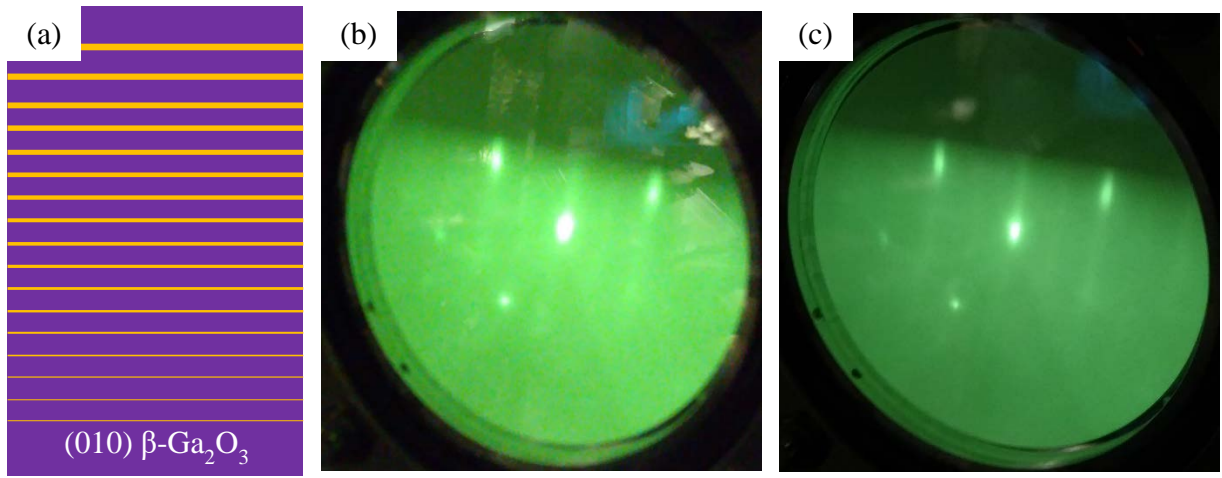

**Figure 2.** (a) Schematic showing the grown SL structure with μ-Fe$_2$O$_3$ (orange) and β-Ga$_2$O$_3$ (purple). RHEED images during β-Ga$_2$O$_3$ (b), and μ-Fe$_2$O$_3$ growth (c).

X-ray diffraction (XRD) is performed using a SmartLab diffractometer in quasi-triple axis mode and a 4-bounce (220) Ge monochromator. Wide angle 2θ/ω scans do not show the presence of either the α or γ phases of Fe$_2$O$_3$. The green and blue vertical lines in Figure 3a mark the expected Bragg angles for the high intensity reflections of the α and γ phases, respectively. No measurable intensity at these angles is observed, suggesting a lack of these additional phases. Nonetheless, we note that this simplistic analysis is not fully conclusive, as those phases, if present, could have some preferential orientation, whereas this basic Bragg angle approach assumes random (powder) orientation.

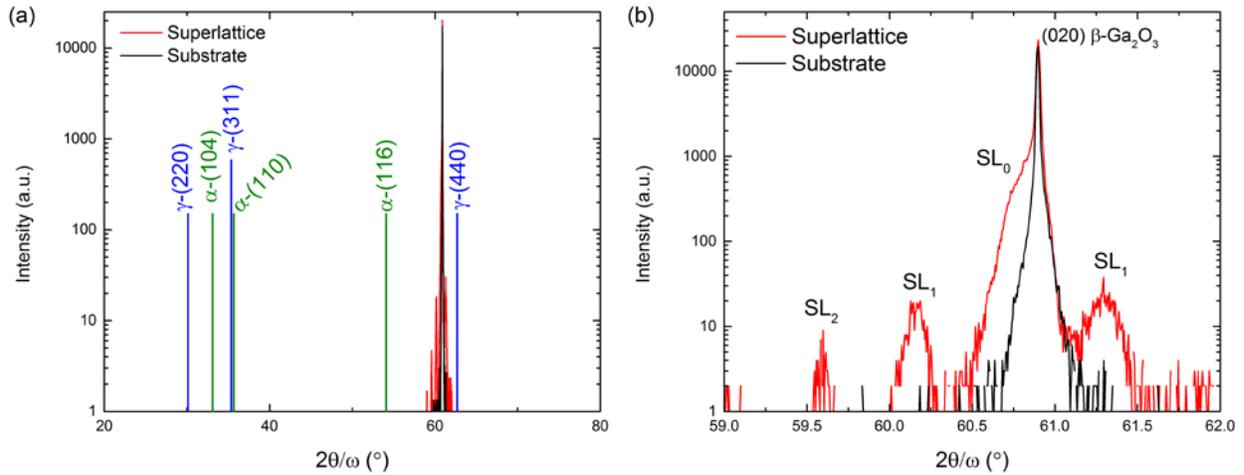

**Figure 3.** 2θ/ω scans of both the (010) β-Ga₂O₃ substrate (black) and the μ-Fe₂O₃/β-Ga₂O₃ SL structure (red). (a) Wide-angle scan showing no evidence of unintended Fe₂O₃ phases. Markers for three high intensity peaks of both the α (green) and γ (blue) phases are shown, along with their corresponding indices, the relative height of the markers is arbitrary.

High-resolution XRD, shown in Figure 3b, reveals superlattice diffraction peaks associated with the multilayer structure. Since the thickness of each period varies, lower order diffraction peaks from the super structure may overlap. To gain more information from such an aperiodic structure we fit superlattice data from Figure 3b using the open source dynamical X-ray diffraction software, CADEM.[23] The input structure is calculated using the growth rate calibrations, and we treat the substrate as the first layer in the structure with a thickness of 5 μm. For simplicity, we assume that the out-of-plane lattice parameter for each μ-Fe₂O₃ layer is the same and use it as the single free parameter in our fits, shown in Figure 4. We find that the average b lattice parameter is 3.12 ± 0.4 Å. Within the error bars, this measurement spans the predicted lattice constants for unstrained (3.07 Å) and fully strained (3.16 Å) μ-Fe₂O₃.

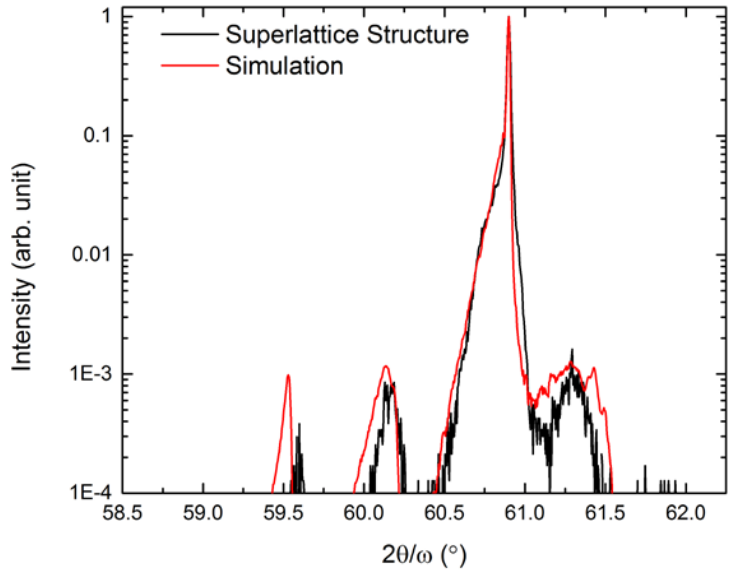

**Figure 4.** Comparison between the XRD data from the superlattice sample (black) and the fit (red) performed using dynamical XRD simulation.

Scanning transmission electron microscopy (STEM) is performed to determine the atomic-scale structural quality of the epilayer using an FEI probe-corrected Titan3 80-300 STEM with an accelerating voltage of 300 kV and a beam current of 80 pA. High angle annular dark field (HAADF) mode images reveal the sample to be crystalline throughout. However, Figure 5a shows that for periods with $\mu$-$Fe_2O_3$ thicknesses greater than 25 ML, the boundaries between the $\mu$-$Fe_2O_3$ and $\beta$-$Ga_2O_3$ layers are no longer continuous. This 25 ML layer marks a region during the growth when the O-plasma source became temporarily unstable resulting in degraded material quality. The layers above the marked 25 ML line in Figure 5a show signs of pitting during the growth. In the Ga rich growth regime, $Ga_2O$ suboxides can form at the surface which dissolve $Ga_2O_3$ forming pits[24]. Nevertheless, atomic resolution images show the same crystal structure throughout (Figure 5b), demonstrating lattice matching of μ-Fe$_2$O$_3$ to β-Ga$_2$O$_3$. The vertical Fe$^{3+}$ rich section in Figure 5b could appear from Fe$^{3+}$ selectively occupying pitted regions during growth.

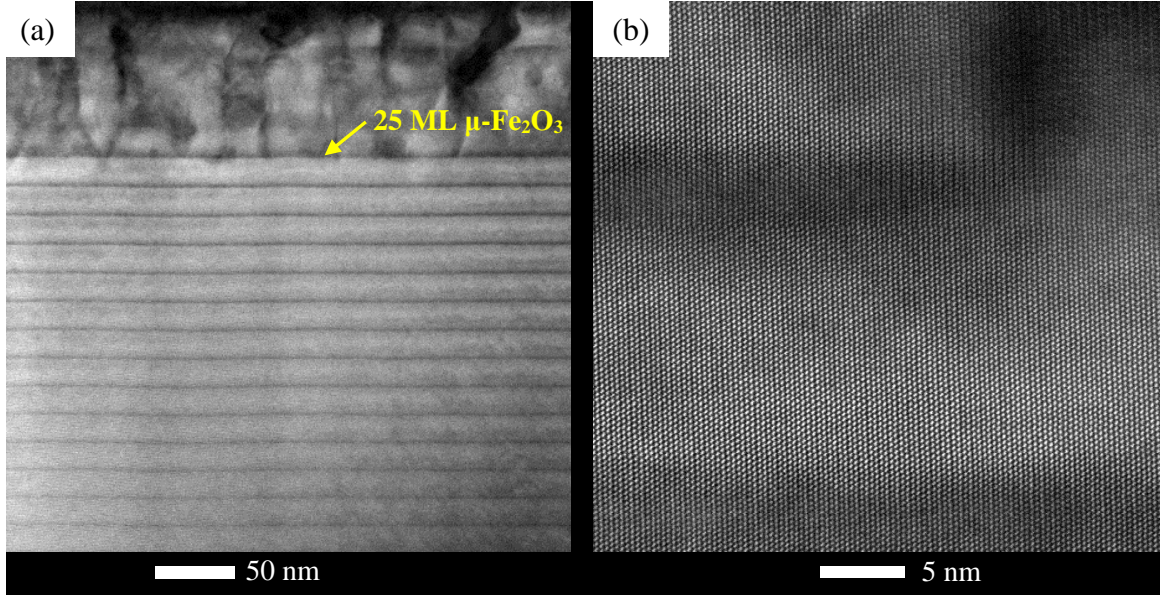

**Figure 5.** HAADF-mode STEM images (a) of the increasing ML superlattice structure. The yellow arrow and label mark the 25 ML μ-Fe$_2$O$_3$ layer. (b) of the registry of the Fe containing regions to the β-Ga$_2$O$_3$ spacer layers. The zone axis is [001].

To investigate the magnetic properties of μ-Fe$_2$O$_3$ SQUID magnetometry is performed on both the SL structure as well as the substrate. The applied field (H) is scanned in a hysteretic fashion from between ±50 kOe along the [102] direction (in-plane) and the induced moment is measured. The diamagnetic background along with the pole hysteresis is removed by subtraction of the substrate magnetization, shown at room temperature in Figure 6a. Normalizing by the volume of the grown μ-Fe$_2$O$_3$ (excluding the Ga$_2$O$_3$), yields the magnetization of the μ-Fe$_2$O$_3$ layers, Figure 6b. Ferromagnetic hysteresis is observed across the measured temperature range (10 K to 300 K). Additionally, a clear paramagnetic component is also observed. Both the coercive field (H$_c$) of the ferromagnetic component and the magnitude of the non-hysteretic paramagnetic component increase with decreasing temperature. Evidence of a Morin transition, which occurs in α-Fe$_2$O$_3$ at 260 K corresponding to a transition from antiferromagnetism to weak-ferrimagnetism, is not observed. In 1960, Geller hypothesized that a structure of $Fe_2O_3$ isomorphic to that of $\beta$-$Ga_2O_3$ would be antiferromagnetic, in agreement with the DFT results.[25] However, we clearly observe the presence of either ferro- or ferrimagnetism.

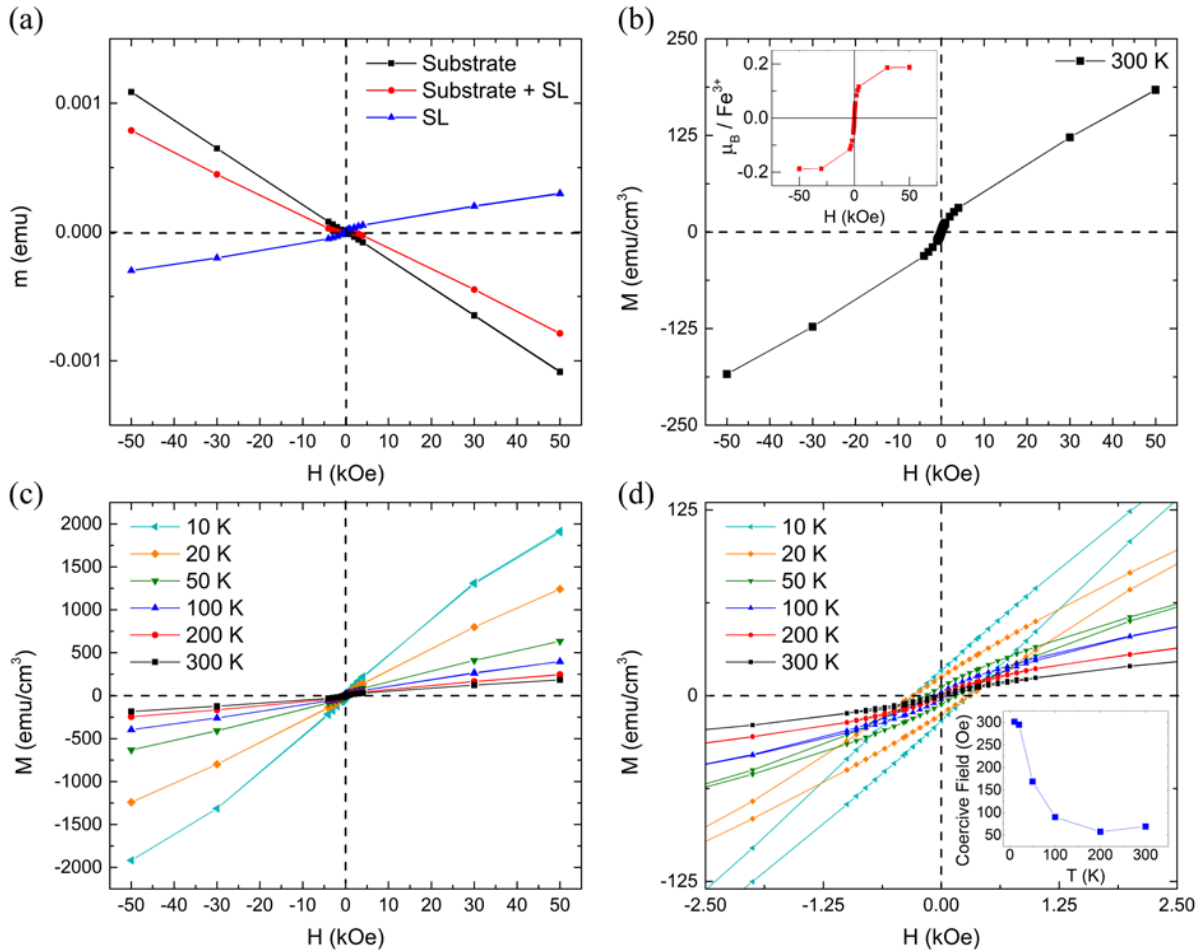

**Figure 6.** (a) SQUID magnetometry showing the raw magnetic hysteresis from the substrate and SL structure as well as the subtraction. (b) Magnetization hysteresis of the SL at 300 K. Inset: The ferromagnetic hysteresis normalized by the $Fe^{3+}$ content. (c,d) Magnetization hysteresis of the SL measured at different temperatures. Inset: Coercive field ($H_c$) versus temperature.

One possible explanation for the unexpected occurrence of the observed paramagnetic and ferromagnetic responses is that our current growth conditions do not fully suppress inter diffusion of $Fe^{3+}$ and $Ga^{3+}$ ions within the layers, and that tetrahedral and octahedral sites at the $Fe_2O_3$/ $Ga_2O_3$ interface may have different preferences for Fe and Ga ions, leading to mixed-cation interface layers with ferromagnetic coupling.

To examine this hypothesis, we first explore the energetics and magnetism for site preferences in $\beta$-$Ga_2O_3$ and $\mu$-$Fe_2O_3$ with DFT calculations by using $1\times4\times2$ supercells with analogous settings described above. As a reference level, we take the $GaO_{3/2}$ and $FeO_{3/2}$ energies for non-defective bulk material, as is the customary assumption for dilute semiconductors.[26] The site-dependent energetic costs for substitution of $Ga^{3+}$ into $\mu$-$Fe_2O_3$ and $Fe^{3+}$ into $\beta$-$Ga_2O_3$ in the previous dominant identified neutral charge state[12] are summarized in Table 2. For substitution of $Ga^{3+}$ into $\mu$-$Fe_2O_3$ there is a 0.31 eV preference for tetrahedral occupancy. For $Fe^{3+}$ in $\beta$-$Ga_2O_3$ the situation is less clear. While the octahedral site always relaxes to the $1\mu_B$ spin state, the tetrahedral site can be relaxed to spin states with 1, 3, and $5\mu_B$, whose energy relative to the octahedral site are +1.02, +0.28, and -1.04 eV, respectively. The energy of the $3\mu_B$, spin state is similar to the 0.31 eV calculated recently in[12] using HSE, where however only one spin state was examined. In our calculations, $Ga^{3+}$ within μ-$Fe_2O_3$ (in either site) results in a magnetic moment of 5 $\mu_B$. The moment on $Ga^{3+}$ is 0, all other $Fe^{3+}$ stay at ±4 $\mu_B$, leaving one of the moments uncompensated. The remaining 1 $\mu_B$ results from uncompensated partial moments on the O atoms. For nearest and second-nearest neighbor positions, $Fe^{3+}$ cations are calculated to couple antiferromagnetically, while for larger distances, the exchange interaction vanishes, suggesting a paramagnetic response for the dilute case.

**TABLE 2**. Magnetic moment (m) and calculated energy costs for $Ga^{3+}$ or $Fe^{3+}$ substitutional solute ions in $\mu$-$Fe_2O_3$ or $\beta$-$Ga_2O_3$ for both tetrahedral ($\Delta E_{tet}$) and octahedral ($\Delta E_{oct}$) site occupancy.

| Solute | Solvent | m ($\mu_B$) | $\Delta E_{tet}$ (eV) | $\Delta E_{oct}$ (eV) | $\Delta E_{tet-oct}$ (eV) |
|--------|---------|-------------|-----------------------|-----------------------|---------------------------|
| $Ga^{3+}$ | $\mu$-$Fe_2O_3$ | 5 | 0.65 | 0.96 | -0.31 |
| $Fe^{3+}$ | $\beta$-$Ga_2O_3$ | 1 | 2.73 | 1.71 | 1.02 |
| $Fe^{3+}$ | $\beta$-$Ga_2O_3$ | 3 | 1.99 | – | 0.28 |
| $Fe^{3+}$ | $\beta$-$Ga_2O_3$ | 5 | 0.67 | – | -1.04 |

In support of this potential explanation for the paramagnetic component, HAADF-mode (Z-contrast) STEM imaging in Figure 5b shows that the interfaces may indeed not be perfectly sharp from a chemical standpoint, as there is a clear contrast gradient between the layers. To investigate this further, energy dispersive x-ray (EDX) mapping is performed separately from the images in Figure 5 on an FEI Image Corrected Titan3™ using a four-quadrant SuperX detector. The EDX maps are consistent with the high-quality growth observed in the HAADF image (Figure 7). Confirmation of the short length scale composition variation at the interface regions, discussed below, requires however higher resolution EDX mapping than what is presented in Figure 7 and is the subject of on-going studies.

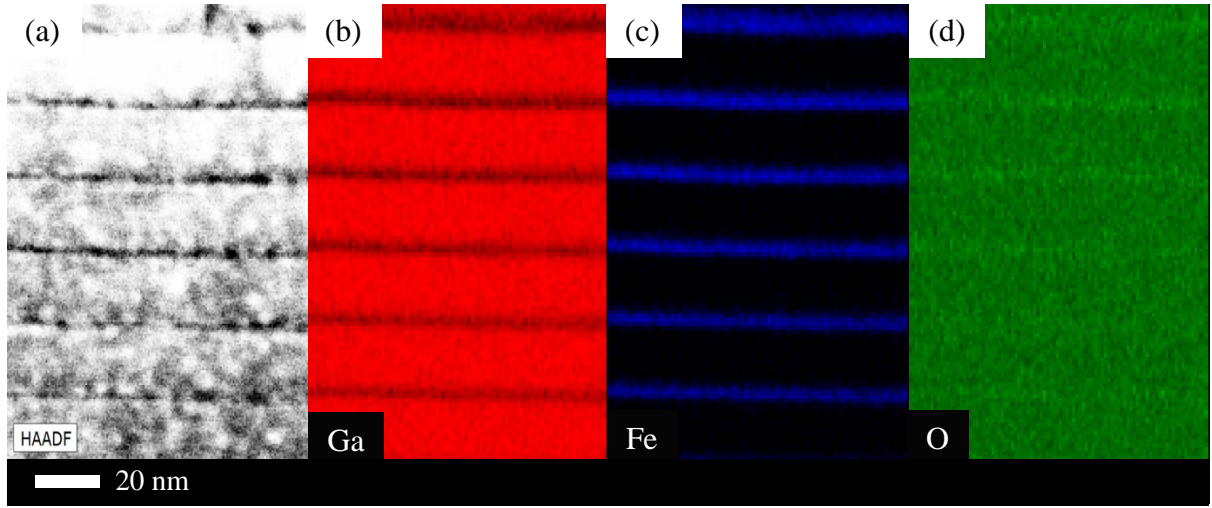

**Figure 7.** (Color online) (a) High angle annular dark field STEM image of a portion of the SL structure. EDX mapping of the image in (a) showing the distribution of Ga (b), Fe (c) and O (d).

The situation at the interface is examined by calculations for a 1×8×1 heterostructure with 1×4×1 β-$Ga_2O_3$ and μ-$Fe_2O_3$ halves. After examining all possible permutations for mixed interface layers while maintaining stoichiometry, we find the lowest energy for interface layers occurs when either all tetrahedral or all octahedral sites are occupied by $Fe^{3+}$, coupling ferromagnetically with 5 $μ_B$ per interface cation, whereas the other sites are occupied by $Ga^{3+}$. The calculated energy difference between these two configurations is less than 10 meV/nm$^2$, and thus not significant within DFT accuracy. In contrast, flat interfaces as well as interfaces where both $Ga^{3+}$ and $Fe^{3+}$ occupy alternating tetrahedral and octahedral sites are 0.7 eV/nm$^2$ higher in energy, and both result in antiferromagnetic coupling. Thus, mixed interfaces with site separation of Fe and Ga between octahedral and tetrahedral sites is energetically clearly preferable, resulting in ferromagnetic coupling between $Fe^{3+}$ spins in the interface layer.

These calculations suggest that the ferromagnetic response from the heterostructure comes from the interface regions. To test this hypothesis, we first isolate the ferromagnetic component of the magnetization by subtracting the low-field linear paramagnetic contribution in Figure 6(b). This yields a ferromagnetic saturation magnetization of 30 emu/cm$^3$, or 0.2 $\mu_B$ per deposited $Fe^{3+}$ cation as shown in the inset of Figure 6(b). The predicted ferromagnetism requires that the cation content of the ferromagnetic layer be comprised of 50% $Fe^{3+}$, (i.e. the $\mu$-Fe$_2$O$_3$ layers must be thicker than the inter-diffusion length scale, although this length scale is unknown). The deposited structure consists of many $Fe^{3+}$ containing layers of varying thicknesses some of which are less than a few ML. Clearly, $\mu$-Fe$_2$O$_3$ layers that are less than a single ML cannot contribute. For $\mu$-Fe$_2$O$_3$ layers which contribute to the overall ferromagnetic signal there will be two interfacial layers that are ferromagnetic with 2 $Fe^{3+}$ cations per unit cell at 5 $\mu_B$ each, totaling 20 $\mu_B$ per unit cell. Excluding layers whose calibrated growth thickness is less than the minimum combined surface roughness of the two interface regions ($\sim$1.5 nm) leaves 8 separate $\mu$-Fe$_2$O$_3$ layers capable of exhibiting interface ferromagnetism in the entire heterostructure. Normalizing by the total Fe deposition gives $\sim$0.3 $\mu_B$/$Fe^{3+}$ which is in good agreement with the experimental results. However, this simple calculation neglects inter diffusion of $Ga^{3+}$ and $Fe^{3+}$ and therefore should be considered an upper bound. Although there is remarkable agreement between the magnetic measurements and ab-initio prediction, additional samples are needed to more directly examine and isolate the magnetic properties of the $\mu$-Fe$_2$O$_3$ / $\beta$-Ga$_2$O$_3$ interface. This will require optimization of the growth conditions to prevent interdiffusion and enable magnetically phase pure material, the subject of ongoing work.

To summarize, we demonstrate a new polytype of Fe$_2$O$_3$ isomorphic to monoclinic $\beta$-Ga$_2$O$_3$, hereby called $\mu$-Fe$_2$O$_3$. *Ab-initio* calculations show that $\mu$-Fe$_2$O$_3$ is as thermodynamically stable as other natural polytypes of Fe$_2$O$_3$ and predict it to have an antiferromagnetic ground state. Using PAMBE we grew an increasing ML superlattice of $\mu$-Fe$_2$O$_3$/$\beta$-Ga$_2$O$_3$. XRD yields an average **b** lattice constant of $\mu$-$Fe_2O_3$ of 3.12 ± 0.4 Å, in agreement with the DFT prediction. STEM structural and chemical imaging shows registry of the $Fe_2O_3$ to the $\beta$-$Ga_2O_3$ lattice and demonstrates the high-quality epitaxial growth of $\mu$-$Fe_2O_3$/$\beta$-$Ga_2O_3$ heterostructures. SQUID magnetometry measurements show a ferromagnetic hysteresis at room temperature. *Ab-initio* modeling of the interface region between $\mu$-$Fe_2O_3$/$\beta$-$Ga_2O_3$ demonstrates a site preference of $Fe^{3+}$ ions in the tetrahedral positions and predicts an interface ferromagnetic phase that is consistent with the measured magnetization. These results provide a pathway to add magnetic functionality into wide-bandgap power electronics.

**Corresponding Author**

*E-mail: myers.1079@osu.edu.

ACKNOWLEDGMENT

This work was supported by the Center for Emergent Materials at The Ohio State University, an NSF MRSEC (DMR-1420451) and by the Army Research Office MURI (W911NF-14-1-0016). SCC and WW acknowledge funding from the Waste PD Center, an EFRC funded by DOE-BES (DESC0016584).

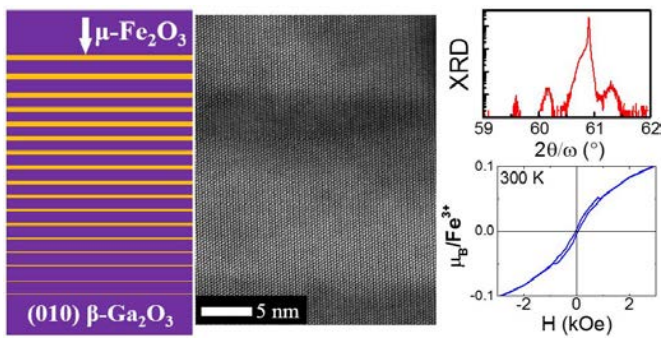

SYNOPSIS

μ-Fe$_2$O$_3$, a new monoclinic form of Fe$_2$O$_3$ is grown on (010)-oriented β-Ga$_2$O$_3$ by molecular beam epitaxy. Atomic-resolution electron microscopy of the μ-Fe$_2$O$_3$/β-Ga$_2$O$_3$ superlattice structure shows an uninterrupted monoclinic lattice throughout. Spontaneous magnetization and magnetic hysteresis persist up to room temperature. Adding magnetic functionality to β-Ga$_2$O$_3$ provides a pathway for high power spintronics.